\documentclass[nofootinbib,aps,preprint]{revtex4}
\usepackage{graphicx}
\begin{document}
\def\be{\begin{equation}}
\def\ee{\end{equation}}
\def\bea{\begin{eqnarray}}
\def\eea{\end{eqnarray}}
\title{Ideal quantum gases: A geometrothermodynamic approach}
\author{Sasha A. Zaldivar$^1$ and Hernando Quevedo$^{1,2}$   }
\email{quevedo@nucleares.unam.mx,sashatf141@gmail.com}
\affiliation{$^1$Instituto de Ciencias Nucleares, 
Universidad Nacional Aut\'onoma de M\'exico, 
 AP 70543, M\'exico, DF 04510, Mexico}
\affiliation{$^2$ 
Dipartimento di Fisica and ICRA, Universit\`a di Roma ``La Sapienza", Piazzale Aldo Moro 5, I-00185 Roma, Italy}

\begin{abstract}
We derive the fundamental thermodynamic equation for Fermi-Dirac and Bose-Einstein quantum gases, which contains the first order contribution due to the quantum nature of the gas particles. Then, we analyze the  fundamental equation in the context of geometrothermodynamics. 
Although the corresponding Hamiltonian does not contain a potential, indicating the lack of classical thermodynamic interaction, we show that the
curvature of the equilibrium space is non-zero and can be interpreted as a measure of the effective quantum interaction between the gas particles. 
In the limiting case of a classical Boltzmann gas, we show that the equilibrium space becomes flat, as expected from the physical viewpoint.  
In addition, we derive a thermodynamic fundamental equation for the Bose-Einstein condensation and, 
using the Ehrenfest scheme, we show that it  can be considered as a first order phase transition which in the equilibrium space corresponds 
to a curvature singularity. This result indicates that the curvature of the equilibrium space can be used to measure in an invariant way the thermodynamic interaction 
in classical and quantum ideal gases.

\end{abstract}

\maketitle


\section{Introduction}
\label{sec:int}

In 1915, Einstein formulated the final version of the gravitational field equations that were based upon 
the astonishing principle ``gravitational interaction = curvature". In this case, the curvature is the Riemannian curvature of the 4-dimensional spacetime.
This principle has been generalized to include all the four interactions known in nature (see, for instance, Ref. \cite{frankel}). 
Indeed, Yang and Mills 
\cite{ym53} demonstrated in 1953 that the field strength (Faraday tensor) of the electromagnetic field can be interpreted as the curvature 
of a principal fiber bundle, with the Minkowski spacetime as the base manifold and the symmetry group $U(1)$
as the standard fiber. Today, it is well known \cite{frankel} that the weak interaction and the strong interaction can be described by the 
curvature of a principal fiber bundle with standard fiber $SU(2)$ and $SU(3)$, respectively. In this sense, we can say that all the field theories 
have a well-formulated geometric description.

On the other hand, Riemannian geometry has been also applied in statistical physics and thermodynamics. To this end, one can consider the 
equilibrium states of the thermodynamic system as points of an abstract space (the equilibrium space). Then, one of the goals of applying differential 
geometry in thermodynamics is to interpret the curvature of the equilibrium space as a measure of the thermodynamic interaction. 
In 1945, Rao \cite{rao45} introduced in the equilibrium space a Riemannian metric 
whose components in local coordinates coincide with the Fisher
information matrix. Rao's original work has been followed up and extended by a large number of
authors (see, e.g., \cite{amari85} for a review). On the other hand, Riemannian geometry in the space of
equilibrium states was introduced by Weinhold \cite{wei75} and Ruppeiner \cite{rup79,rup95},
who defined metric structures as the Hessian of the internal energy and (minus) the entropy, respectively.

Geometrothermodynamics (GTD) \cite{quev07} was proposed recently to take into account the invariance of classical thermodynamics under a change of thermodynamic 
potential \cite{callen}, a property which is not shared by Hessian metrics. In contrast to other geometric approaches, GTD resembles the approach of field theories
in the sense that the symmetry of the theory plays a fundamental role in its geometric description. 
Since different thermodynamic potentials are related by means of Legendre transformations \cite{arnold}, the formalism of GTD makes use of the Riemannian contact structure of the thermodynamic phase space \cite{her73} to handle the interchanges of thermodynamic potentials as coordinate transformations on the phase space. The equilibrium space can then be considered as a particular subspace of the phase space.  As a result,
the formalism of GTD uses as starting point a Legendre invariant metric of the phase space which induces a metric in the equilibrium space whose curvature should describe the thermodynamic interaction. In the case of a classical ideal gas, for which in statistical physics the corresponding Hamiltonian does not contain a potential term that could 
represent any interaction between the gas particles \cite{huang}, 
no thermodynamic interaction is present and one would expect that the corresponding equilibrium space be flat. 
In fact, this physical condition has been used to fix the metric of the phase space \cite{quev07}. The flat equilibrium manifold of a classical ideal gas has been investigated in detail in \cite{qsv15}. In particular, it was found that  there exists a deep relationship between geodesics and quasi-static processes, 
 which gives rise to a relativistic like structure of the equilibrium manifold. 

In the case of ideal quantum gases, as described in statistical physics, the Hamiltonian also does not contain a potential term 
and the gas particles are assumed to have no interaction between them \cite{huang,greiner}. 
This means that from the point of view of the Hamiltonian approach a quantum gas 
is a system without thermodynamic interaction. One would then expect that the corresponding equilibrium manifold is also flat due to the lack of interaction.
Nevertheless, we know that the physical properties of ideal quantum gases are different from those of a classical ideal gas. The question arises whether our 
geometric approach is able to take into account those physical differences in a consistent manner. The main purpose of the present work is to show that in the framework of GTD the ideal quantum gases are represented by non-flat equilibrium manifolds, taking into account the quantum nature of the gas particles. This we show in the case of Fermi-Dirac and Bose-Einstein quantum gases. Moreover, we analyze the limiting case of Bose-Einstein condensation and show that it can be interpreted in the Ehrenfest scheme and in GTD as a first order phase transition.

This paper is organized as follows. In Sec. \ref{sec:gtd}, we review the main aspects of GTD which are necessary in order to take into account the Legendre invariance of 
classical thermodynamics. In Sec. \ref{sec:iqg}, we analyze ideal quantum gases satisfying the Fermi-Dirac and Bose-Einstein statistics and derive the corresponding fundamental equation, 
which is necessary to carry out the geometrothermodynamic approach. In Sec. \ref{sec:ebc}, we investigate the thermodynamic and geometrothermodynamic properties of Bose-Einstein condensates.  Finally, in Sec. \ref{sec:con}, we 
discuss our results.


\section{Review of geometrothermodynamics}
\label{sec:gtd}

The starting point of the GTD formalism is the thermodynamic phase space ${\cal T}$ which is constructed as follows. 
A thermodynamic system with $n$ degrees of freedom is described by a set of $n$ extensive variables, $E^a$, $n$ intensive variables, $I^a$ $(a=1,...,n)$, and a thermodynamic potential, $\Phi$. Let us consider $Z^A=\{\Phi, E^a, I^a\}$ as the coordinates
of the $(2n+1)-$dimensional space ${\cal T}$, where $E^a$ and $I^a$ are interpreted as independent variables. 
The phase space is necessary in GTD in order to treat the Legendre transformations of classical thermodynamics as a change of coordinates. Formally, a Legendre transformation  is defined as \cite{arnold} (we assume throughout Einstein's summation convention for repeated indices)
\begin{eqnarray}
\{Z^A\} \rightarrow \{\tilde{Z}^{A}\} = \{\tilde{\Phi},\tilde{E}^a,\tilde{I}^a\} 
\end{eqnarray}
\be
\Phi = \tilde{\Phi} - \delta_{kl} \tilde{E}^k\tilde{I}^l \, , \quad E^i = -\tilde{I}^i \, , E^j = \tilde{E}^j \, ,\  I^i = \tilde{E}^i \, , \ I^j = \tilde{I}^j\ ,
\label{leg}
\ee
where $i \cup j$ is any disjoint decomposition of the set of indices $\{1, \ldots, n\}$, and $k,l = 1, \ldots, i$. 
In particular, for 
$i=\emptyset$ a Legendre transformation reduces to the identity transformation, for any $j\neq \emptyset$ we obtain partial Legendre transformations, 
and 
for
$i=\{1,...,n\}$, i.e. $j=\emptyset$, Eq.(\ref{leg}) defines a total Legendre
transformation. 

The phase space $\mathcal{T}$ is  endowed with a family of tangent hyperplanes 
(contact structure) defined by the fundamental 1-form $\Theta$ that satisfies the non-integrability condition
\cite{her73} 
\be
\label{gtd.ni}
\Theta \wedge ( d \Theta)^n \neq 0\ .
\ee
According to the Darboux theorem \cite{her73}, 
if we use the coordinates $\{Z^A\}$ of ${\cal T}$, the fundamental 1-form $\Theta$ can be written canonically as
\be
\label{gtd.local}
\Theta =  d \Phi - I_a  d E^a \ , \quad I_a = \delta_{ab} I^b \ .
 \ee 
If we apply a Legendre transformation to $\Theta$, the new 1-form $\tilde{\Theta}$ in coordinates $\{\tilde{Z}^A\}$ reads
\be
\label{gtdlegendre}
\tilde{\Theta} =  d \tilde{\Phi} - \tilde{I}_a  d \tilde{E}^a \ .
\ee
In this sense, it is said that the contact 1-form $\Theta$ is invariant with respect to Legendre transformations. 

In classical thermodynamics, a thermodynamic system is  described equivalently either by a set of equations of state $I^a = I^a(E^b)$ or by a fundamental equation 
$\Phi=\Phi(E^a)$.  
In GTD, it is convenient to use the second option and we assume that for any thermodynamic system the fundamental equation is known.  
Furthermore, the specification of a particular system is realized in GTD by an embedding $\varphi$ of an $n$-dimensional submanifold $\mathcal{E}\in {\cal T}$ into the phase space $\mathcal{T}$ given by
\be
	\label{gtd.emb}
	\varphi:\mathcal{E}\longrightarrow\mathcal{T},
	\ee
or, in coordinates,
\be
\varphi:\{E^a\} \longrightarrow \{\Phi(E^a), E^a, I^b(E^a)\}\ .
\ee

Accordingly, the space $\mathcal{E}$  is determined through the embedding (\ref{gtd.emb}), which is equivalent to specifying the fundamental equation of the system $\Phi(E^a)$. To complete the GTD scheme, it is necessary to incorporate the relations of standard equilibrium thermodynamics into the definition of the space ${\cal E}$. This can be done by demanding that the 
embedding (\ref{gtd.emb}) satisfies the condition
	\be
	\label{gtd.01}
	\varphi^*(\Theta) = 0\ ,
	\ee
where $\varphi^*$ is the pullback of $\varphi$. In coordinates, it takes the form
	\be
	\label{gtd.02}
	\varphi^*(\Theta) = \varphi^*\left( d \Phi - I_a  d E^a\right) = \left(\frac{\partial \Phi}{\partial E^a} - I_a\right) d E^a  = 0.
	\ee
Then, it  follows that
	\be
	\label{gtd.03}
	d\Phi = I_a d E^a , \quad  \frac{\partial \Phi}{\partial E^a} = I_a\ ,
	\ee
which is the first law of thermodynamics. The manifold $\mathcal{E}$  is called the equilibrium space. Moreover, it is demanded that the fundamental equation $\Phi=\Phi(E^a)$ satisfies all the remaining laws of equilibrium thermodynamics. 

The next step to construct the formalism of GTD consists in equipping the phase space with a metric $G$, which must be invariant with respect to Legendre transformations. This is necessary in order to incorporate the Legendre invariance of classical thermodynamics in the entire geometric structure of GTD. 
Consequently, the triad 
$(\mathcal{T},\Theta, G)$ becomes a Riemannian contact manifold that is also Legendre invariant. 
It turns out that Legendre invariant metrics $G=G_{AB}dZ^A dZ^B$ in ${\cal T}$ can be split into three different classes that 
can be written as 
\be
G^{^{I}}=  (d\Phi - I_a d E^a)^2 + (\xi_{ab} E^a I^b) (\delta_{cd} dE^c dI^d) \ ,
\label{GI}
\ee
\be 
G^{^{II}}= (d\Phi - I_a d E^a)^2 + (\xi_{ab} E^a I^b) (\eta_{cd} dE^c dI^d) \ ,
\label{GII}
\ee
\be	
\label{GIII}
G^{{III}}  =(d\Phi - I_a d E^a)^2  + \sum_{a=1}^n E_a I_a   d E^a   d I^a \ ,
\ee
where $\xi_{ab}$ is a diagonal constant $(n\times n)$-matrix, $\delta_{ab}= {\rm diag}(1,1,\cdots,1)$, 
and $\eta_{ab}= {\rm diag}(-1,1,\cdots,1)$.
Moreover, $G^I$ and $G^{II}$ are invariant with respect to total Legendre transformations whereas $G^{III}$ is also invariant under partial transformations. 

One interesting feature of the formalism of GTD is that the equilibrium space ${\cal E}$ can be endowed with a metric $g$ in a canonical way, namely, by means of the pullback $\varphi^*(G)=g$ so that $g$ inherits the properties of $G$.  
 Then, from Eqs.(\ref{GI}), (\ref{GII}), and (\ref{GIII}), we obtain
\be
g^{{I}}_{ab} =   \beta_\Phi \Phi  \delta_a^{\ c}
\frac{\partial^2\Phi}{\partial E^b \partial E^c}   ,
\label{gdownI}
\ee
\be
g^{II}_{ab} =   \beta_\Phi \Phi  \eta_a^{\ c}
\frac{\partial^2\Phi}{\partial E^b \partial E^c}   ,
\label{gdownII}
\ee
\be
g^{{III}} = \sum_{a=1} ^n \left(\delta_{ad} E^d \frac{\partial\Phi}{\partial E^a}\right) \delta^{ab} \frac{\partial ^2 \Phi}{\partial E^b \partial E^c}
dE^a dE^c \ ,
\label{gdownIII}
\ee
respectively, where $\delta_a^{\ c}={\rm diag}(1,\cdots,1)$, $\eta_a^{\ c}={\rm diag}(-1,1,\cdots,1)$. The constant $\beta_\Phi$ does not affect the geometric properties of $g$, and represents the degree of homogeneity of the thermodynamic potential $\Phi$ \cite{qqs19}. In obtaining the above metrics we have used the first law of thermodynamics (\ref{gtd.03}) and 
the Euler identity  
\be
\beta_{ab} I^aE^b = \beta_\Phi \Phi\ ,
\label{euler}
\ee
where the coefficients $\beta_{ab}=\xi_{ab}={\rm diag}(\beta_1,\beta_2,...,\beta_n)$ are determined by the degrees of (quasi)homogeneity of the variables $E^a$, entering the fundamental equation $\Phi=\Phi(E^a)$, i.e., the constants  $\beta_a$ that satisfy the quasi-homogeneity condition $\Phi(\lambda^{\beta_a} E^ a)= \lambda^{\beta_\Phi}\Phi(E^a)$.

For later use consider the case of a system with two degrees of freedom, $n=2$. Then, $\Phi=\Phi(E^1,E^2)$ and  
\be
\label{gI2D} 
g^{I}= \Phi \left[\Phi_{,11} (d E^1)^2 + 2 \Phi_{,12} dE^1 dE^2 + \Phi_{,22} (dE^2)^2\right]\,,
\ee
\be
g^{II}= \Phi \left[-\Phi_{,11} (d E^1)^2  + \Phi_{,22} (dE^2)^2\right]\,,
\label{gII2D}
\ee
\be
g^{III}= E^1 \Phi_{,1} \Phi_{,11} (dE^1)^2 +
\left(E^1\Phi_{,1} + E^2\Phi_{,2}\right) \Phi_{,12} dE^1 dE^2
+ E^2 \Phi_{,2} \Phi_{,22} (dE^2)^2\ ,
\label{gIII2D}
\ee
where $\phi_{,a} = \frac{\partial \phi}{\partial E^a}$, and we have set $\beta_\Phi=1$, without loss of generality. 
One of the goals of GTD is to represent phase transitions as singularities of the equilibrium space. The idea is simple. During a phase transition the laws of equilibrium thermodynamics break down and non-equilibrium effects appear that cannot be considered with the standard formalism. A curvature singularity represents a location where the formalism of classical differential geometry breaks down. Therefore, we expect intuitively a connection between curvature singularities and phase transitions. This has been shown explicitly in the case of 2-dimensional  GTD by using Euler's identity (\ref{euler}). In fact,  it turns out that the curvature singularities of the  above metrics are determined by the conditions \cite{qqs22a,qqs22b}
\bea
I: && \qquad \Phi_{,11}\Phi_{,22} -(\Phi_{,12})^2
=0 \ ,\label{singirev} \\
II: && \qquad \Phi_{,11} \Phi_{,22} 
=0\ , \label{singiirev}  \\
III: && \qquad \Phi_{,12}= 0 \ . \label{singiiirev} 
\eea

Condition $I$ corresponds to the stability condition in classical thermodynamics, and is usually associated with first order phase transitions  \cite{callen}. Conditions $II$ and $III$ involve the second order derivatives of the thermodynamic potential $\Phi$ and usually are associated with second order phase transitions, according to Ehrenfest's scheme \cite{callen}.


\section{Fermi-Dirac and Bose-Einstein ideal quantum gases}
\label{sec:iqg}

Let us consider an ideal gas consisting of $N$ non-interacting identical particles, with mass $m$ and momentum $p_i$, inside a 3-dimensional box of volume $V$. 
According to the standard approach of statistical physics, this system can be described by the Hamiltonian \cite{huang}
\be
{\cal H} = \sum_{i=1}^N \frac{p_i^2}{2m}\ .
\label{ham}
\ee
If we take into account the physical nature of the particles with a maximum occupation number of one or infinity, the ideal gas can be either a Fermi gas, a Bose gas or a Boltzmann gas. In the first two cases, the quantum nature of the particles (fermions or bosons) is the main characteristic of the system, whereas the Boltzmann gas is composed of classical identical particles. From a statistical point of view, the fact that no potential is present in the Hamiltonian (\ref{ham}) indicates the lack of thermodynamic interaction.  

The purpose of this section is to express the thermodynamic properties of the ideal quantum gases in such a representation that it can be used in the context of GTD. 
Although there are several possibilities to derive the statistical model from which the thermodynamic limit could be computed, in the case of spin-less quantum particles, it is convenient to use the grand partition function \cite{huang}
\be
{\cal Q}(V,T,\mu) = \prod_{p}\left[ 1 + \epsilon e^{\beta(\mu-\varepsilon_p)}\right]^\epsilon ,\quad \beta= \frac{1}{k_{_B}T}\ , \quad \varepsilon_p = \frac{p^2}{2m}\ ,
\quad \epsilon=\pm 1
\ee
where $k_{_B}$ is the Boltzmann constant, $T$ is the temperature, $V$ the volume, and $\mu$ represents the chemical potential. The constant $\epsilon$ indicates the type of gas, with $\epsilon=+1$ for a Fermi gas and $\epsilon = -1$ for a Bose gas. The main thermodynamic quantities can be obtained in a straightforward manner 
from the grand partition function as
\be
U=-\frac{\partial}{\partial \beta} \ln {\cal Q}\ , \quad N= \frac{1}{\beta}\frac{\partial}{\partial \mu} \ln {\cal Q} \ ,
\ee
where $U$ is the internal energy and $N$ is the particle number. In the case of quantum gases, we obtain
\be
U= \sum_p \frac{\varepsilon_p}{e^{\beta(\varepsilon_p-\mu)}+\epsilon}\ ,\quad
N= \sum_p \frac{1}{e^{\beta(\varepsilon_p-\mu)}+\epsilon}\ . 
\label{unstat}
\ee
Notice that in this statistical representation the difference between Bose and Fermi gases is formally contained in the parameter $\epsilon$ only. We will see 
that this particularity holds also in the thermodynamic limit.       

\subsection{The fundamental equation}
\label{sec:feq}

There are several equivalent ways to find the fundamental equation of the ideal quantum gases in the thermodynamic limit. In order to compare our results with the well-known Sackur-Tetrode equation for the entropy of a classical ideal gas, we choose the entropy $S=S(U,V,N)$ as the thermodynamic potential that determines the fundamental equation. To find the entropy it is convenient to use the free energy which is related to the entropy through the Legendre transformation $F=U-TS$. Moreover, for the free energy we can use the expression $F= - PV + \mu N$ together with the standard equations of state $PV= Nk_{_B} T$ and $U=\frac{3}{2} N k_{_B} T$ so that we finally obtain 
\be
S= \frac{1}{T}\left(\frac{5}{3} U - \mu N\right) \ .
\label{ent}
\ee
To proceed with the evaluation of this equation we need the expressions for $U$ and $N$ in the appropriate limit.

In the thermodynamic limit $V\rightarrow \infty$, the possible values of $p$ represent a continuum, and so we can replace the sum over all values of $p$ by an integral, i. e., \cite{huang}
\be
\sum_{p} \rightarrow \frac{4\pi V}{h^3} \int p^2 dp = \frac{2\pi V}{h^3} (2m)^{3/2} \int \varepsilon^{1/2} d\varepsilon  \ .
\ee
Then, the energy and particle number (\ref{unstat}) become
\be
U = \frac{3}{2}  k_{_B}  T V \left(\frac{mk_{_B} T}{2\pi \hbar^2}\right)^{3/2} h_{\frac{5}{2}}^\epsilon (z)  \ ,\quad
N = V  \left(\frac{mk_{_B} T}{2\pi \hbar^2}\right)^{3/2} h_{\frac{3}{2}}^\epsilon (z)  \ ,
\label{unh}
\ee
respectively, where we have introduced the new variable $x=\beta\varepsilon$ and $z=e^{\beta\mu}$ is the fugacity. Moreover, we have introduced the notation 
\be
h^\epsilon_n(z) = \frac{1}{\Gamma(n)} \int_0^\infty \frac{x^{n-1}}{z^{-1} e^x + \epsilon} dx 
\ee
for the integrals that appear in the expressions for $U$ and $N$. In the literature, it is common to use the notations $h_n^+(z)=f_n(z)$ and $h_n^-(z)=g_n(z)$ which are known as the Fermi and Bose integrals, respectively. The index $n$ is known as the integral order which depends on the dimension of the system as follows 
from Eq.(\ref{unh}).

The integral $h_n^\epsilon(z)$ for $z<<1$  can be represented as the series
\be
 h_n^\epsilon(z) = \sum_{j=1}^\infty (-\epsilon)^{j+1} \frac{z^j}{j^n} 
\ee
which is useful for concrete calculations.  In fact, the limit of small fugacity is considered as the classical limit of quantum gases in which we obtain from 
Eq.(\ref{unh})
\be
U=\frac{3}{2}  k_{_B}  T V \left(\frac{mk_{_B} T}{2\pi \hbar^2}\right)^{3/2}  \left(z-\epsilon\frac{z^2}{2^{5/2}} \right) \ ,\quad
N=  V  \left(\frac{mk_{_B} T}{2\pi \hbar^2}\right)^{3/2}  \left(z - \epsilon\frac{z^2}{2^{3/2}} \right) \ ,
\label{untd}
\ee
where only quadratic terms in $z$ have been taken into account. We now use the expression for $N$ to express the fugacity in terms of $N/V$. To this end, we replace the truncated series $z=z_1 (N/V) + z_2 (N/V)^2$  into the expression for $N$ and compare the terms on both sides of the equation in such a way that the constants $z_1$ and $z_2$ can be determined. In this manner, we obtain
\be
z = \frac{N}{V} \left(\frac{2\pi \hbar^2}{mk_{_B} T}\right)^{3/2}   +\epsilon \frac{N^2}{2^{3/2} V^2} \left( \frac{2\pi \hbar^2}{mk_{_B} T} \right)^{3}\ .
\label{fuga}
\ee
Notice that the condition of the classical limit $z<<1$ implies that 
\be
\frac{N}{V} \left(\frac{2\pi \hbar^2}{mk_{_B} T}\right)^{3/2}   << 1 \ .
\label{cond}
\ee
  
The expression (\ref{fuga}) for the fugacity can now be used to eliminate $z$ from the internal energy $U$ and to evaluate the chemical potential $\mu =k_{_B} T \ln z$, which can then be replaced into the expression for the entropy (\ref{ent}). Then, we obtain
\be
S = \frac{5}{2} N k_{_B} - N k_{_B} \ln\left[\frac{N}{V} \left( \frac{2\pi \hbar^2}{mk_{_B} T} \right)^{3/2}\right] + \epsilon \frac{N^2 k_{_B}}{2^{7/2}V} 
\left( \frac{2\pi \hbar^2}{mk_{_B} T} \right)^{3/2}  \ ,
\label{feq}
\ee 
where we have considered only the leading terms in the limit of large temperature. This is the fundamental equation for ideal quantum gases. Notice that we are using the temperature $T$ instead of the internal energy $U$ in the fundamental equation (\ref{feq}). One can, of course, use the corresponding equation of state in order to replace $T$ by $U$ so that the entropy will depend on extensive variables only. However, for the geometric analysis we will perform in the following section, the use of $T$ or $U$ is not relevant because the  equation of state that relates $T$ and $U$ can be considered as a diffeomorphism, which does not affect the geometric properties of the underlying manifold. We will use the temperature as thermodynamic variable because it allows us to easily handle the physical limits of the fundamental equation 
(\ref{feq}).  

The Boltzmann limit of the fundamental equation (\ref{feq}) corresponds to the limit of high temperature ($T\rightarrow \infty$), i.e.,
\be
S = \frac{5}{2} N k_{_B} - N k_{_B} \ln\left[\frac{N}{V} \left( \frac{2\pi \hbar^2}{mk_{_B} T} \right)^{3/2}\right] 
\ ,
\label{feq1}
\ee    
which is equivalent to the Sackur-Tetrode equation for the classical ideal gas. 

The fundamental equation (\ref{feq}) for ideal quantum gases indicates that the new term proportional to $\epsilon$ is the result of the quantum nature of the system. So, it can be interpreted as the term responsible for the thermodynamic quantum interaction and, for large values of the temperature, it represents a perturbation of the classical Boltzmann gas. In this sense, from a thermodynamic point of view, we can consider a quantum gas as a perturbation of a classical gas. 
This interpretation is consistent with the virial expansion approach of quantum gases as presented, for instance, in \cite{huang}. 
We will show in the next section that the geometrothermodynamic approach reinforces this interpretation.


\subsection{Geometrothermodynamic analysis}
\label{sec:gtd1}

According to the GTD approach presented in Sec. \ref{sec:gtd}, to find  the metric of the equilibrium manifold it is enough to have the explicit expression of the fundamental equation. This has been done in the previous section. To calculate the metric it is convenient to choose geometric units with $\hbar=1=k_{_B}$. In addition, we can choose $m=2\pi$ without loss of generality. Then, the fundamental equation (\ref{feq}) can be expressed as
\be
S= S_0 - N \ln \left( \frac{N}{V T^{3/2}}  \right) + \epsilon \frac{b N^2}{ V T^{3/2}}    \ ,
\label{feq2}
\ee
where $b=1/2^{7/2}$, and  $S_0$ can be chosen as an additive constant under the condition that the total number of particles $N$ is a constant. This means that we will consider an isolated system with a fixed particle number, and the entropy depends explicitly  on $V$ and $T$ only.

To calculate the metrics (\ref{gI2D})-(\ref{gIII2D}) of the equilibrium manifold ${\cal E}$ with the above fundamental equation we must identify $S$ with the thermodynamic potential $\Phi$ and the coordinates of ${\cal E}$ as $E^a = (T,V)$. Then, we obtain 
\be
g^I = SN\left[\frac{3}{2}\left(-1+\frac{5}{2}\frac{\epsilon b N}{VT^{3/2}}\right) \frac{dT^2}{T^2} + 3\frac{\epsilon b N}{V^2 T^{5/2}} dV dT + \left(-1+\frac{2\epsilon b N}{VT^{3/2}}\right) \frac{dV^2}{V^2}\right] ,
\ee
\be
g^{II}= 
SN\left[-\frac{3}{2}\left(-1+\frac{5}{2}\frac{\epsilon b N}{VT^{3/2}}\right) \frac{dT^2}{T^2} + \left(-1+\frac{2\epsilon b N}{VT^{3/2}}\right) \frac{dV^2}{V^2}\right] ,
\ee
\be
g^{III}= {N^2}\left( \frac{\epsilon b N}{VT^{3/2}}  - 1 \right) \left[ \frac{9}{4}\left( 1 - \frac{5}{2}  \frac{\epsilon b N}{VT^{3/2}}  \right) \frac{dT^2}{T^2}
- \frac{15}{4} \frac{\epsilon b N}{V^2T^{5/2}} dT dV  
+ \left( 1 - 2 \frac{\epsilon b N}{VT^{3/2}}  \right) \frac{dV^2}{V^2} \right] \ .
\label{metqig}
\ee
The signature of these metrics is not fixed, but depends on the choice of $\epsilon$, indicating that the quantum nature of the particles can drastically change the 
geometric properties of the corresponding equilibrium manifold. The determinant of the metric can also be zero for certain combinations of the values of  $V$ and $T$.  This can be interpreted as a violation of the second law of thermodynamics at which the thermodynamic and geometrothermodynamic approaches break down \cite{qsv15}. 

It is straightforward to show that in general the curvature of all three metrics is different from zero,
indicating the presence of non-trivial thermodynamic interaction.  Moreover, a computation of the conditions for the existence of curvature singularities (\ref{singirev})-(\ref{singiiirev}) leads to
\bea 
I: &&  7\frac{\epsilon^2b^2N^2}{V^2T^3} - 9 \frac{\epsilon N}{VT^{3/2}} + 2 =0\ ,\\
II : &&  \left(-1+\frac{5}{2}\frac{\epsilon b N}{VT^{3/2}}\right)
 \left(-1+\frac{2\epsilon b N}{VT^{3/2}}\right) = 0\ , \\
III: && \frac{\epsilon b N^2}{V^2T^{5/2}} = 0\ .
\eea
Only the conditions $I$ and $II$ have non-trivial solutions, namely, 
\be
\frac{bN}{V T^{3/2}} =1, \frac{2}{7}, \frac{2}{5}, \frac{1}{2}\ .
\ee
However, considering the value of the constant $b=1/2^{7/2}$, it can be shown that all the roots are for values within the range  
\be
\frac{N}{V} \left(\frac{2\pi \hbar^2}{mk_{_B} T}\right)^{3/2} = \frac{N}{VT^{3/2}}   > 1 \ ,
\ee
a range that obviously contradicts the condition of the classical limit (\ref{cond}) which was assumed to determine the fundamental equation (\ref{feq2}).  We conclude that all the curvature singularities are non physical. Accordingly, there are no phase transitions in the thermodynamic limit of quantum ideal gases, a result that is in agreement with the one obtained in classical thermodynamics \cite{huang}.

We now consider the limit of the Boltzmann ideal gas which is described by the fundamental equation (\ref{feq1}). Using geometric units and $m=2\pi$, we obtain
\be
S= S_0 -\ln\left(\frac{N}{VT^{3/2}}\right) \ .
\ee
If we now consider, for instance, the metric (\ref{metqig}) of the equilibrium manifold reduces to
\be
g^{III}= -N^2\left(\frac{9}{4} \frac{dT^2}{T^2}+ \frac{dV^2}{V^2}\right)\ .
\label{metig}
\ee
It is then straightforward to show that the corresponding curvature tensor vanishes identically. This can also be seen by introducing the new coordinates 
$d\eta = \frac{3dT}{2T}$ and $d\xi = \frac{dV}{V}$  so that the metric (\ref{metig}) acquires a Cartesian like structure, i.e., $ g^{III}= -d\eta^2- d\xi^2$ whose 
curvature is obviously zero.

In GTD, the invariant curvature is interpreted as a manifestation of the intrinsic thermodynamic interaction between the particles of the system. We have shown that the curvature is zero for the Boltzmann gas and non-zero for the Fermi and Bose quantum gases. In the case of the Boltzmann gas, this in agreement with the statistical approach since the corresponding Hamiltonian (\ref{ham}) has no potential term. In the case of the Fermi and Bose ideal gases, however, the potential term is still zero, but the curvature does not vanish, indicating the presence of thermodynamic interaction.   On the other hand, from a physical point of view one expects a detectable difference between classical and quantum ideal gases. We have shown here that the curvature of the equilibrium space is able to take into account this difference. The non-zero curvature represents an ``effective" thermodynamic interaction which is generated by the quantum nature of the gas particles.


\section{Bose-Einstein condensation}
\label{sec:ebc}

According to Eq.(\ref{unh}), the total number of particles of a Bose-Einstein ideal gas
\be
N= V\left( \frac{mk_{_B}T}{2\pi\hbar^2}\right)^{3/2} g_{\frac{3}{2}}(z) \ , \quad z = e^{\frac{\mu}{k_{_B}T}}\ .
\label{eqN}
\ee
depends on the temperature. Since the Bose integral $g_{\frac{3}{2}}(z)$ has its maximum value at $z=1$, the maximum particle number at a fixed 
value of the temperature, $T=T_c$, is reached for $\mu\rightarrow 0$, and is given by the expression
\be
N_{max}= V \left( \frac{mk_{_B}T_c}{2\pi\hbar^2}\right)^{3/2}\zeta  \ .
\ee
where $\zeta = g_{\frac{3}{2}}(1) \approx 2.61$.  
In the same manner, for a given particle number $N$ we can define the critical temperature $T_c$ as
 \cite{huang}
\be
T_c = \frac{2\pi \hbar^2}{m k_{_B}} \left(\frac{N}{V \zeta  }\right)^{2/3} \ ,
\ee
so that $N_{max}$ is reached for $T<T_c$ and $\mu\rightarrow 0$. It turns out that in this limit, a finite number of particles occupies the state with minimum energy 
$\varepsilon =0$. This phenomenom is known as the Bose-Einstein condensation. This particular state can be considered as a mixture of two different phases, one phase contains all the particles with $\varepsilon =0$ and the second phase with $\varepsilon \neq 0$. It is in this sense that the Bose-Einstein condensation can be considered as a phase transition; however, the order of the transition has  not been fixed definitely. In fact, some authors \cite{greiner} argue that it is a second  order transition \cite{vic1}, but others 
\cite{huang} advocate for a first order phase transition by using a different approach. In this work, we will consider the Ehrenfest criterium to determine the order of a phase transition. 

\subsection{Geometrothermodynamic analysis}

To establish the fundamental equation that governs the transition of a bosonic gas into a Bose-Einstein condensate, we must perform a different analysis. In fact, to find out the properties of the bosonic gas, in the last section we investigated the classical limit in which the fugacity is an infinitesimal quantity, $z<<1$. The condensation, instead, corresponds to small values of the chemical potential, i.e., $z\rightarrow 1$. Furthermore, it is not an easy task to handle this case analytically for particular thermodynamic potentials, like the entropy, in such a way that the Ehrenfest scheme can be applied. Indeed, following the standard procedures of statistical physics, it can be shown that the grand potential for a bosonic gas can be written as \cite{huang}
\be
\Xi = - T^{5/2} V g_{\frac{5}{2}}\left(e^{\frac{\mu}{T}}\right) 
\label{gpb}
\ee
where for simplicity we set $k_B=1$, $\hbar =1$, and $m=2\pi$. However, the difficulty of using this grand potential is that it does not lead to the appearance of condensation. To fix this problem, it is necessary to add the grand potential of the ground state
\be
\Xi_0 = T \ln\left(1- e^{\frac{\mu}{T}}\right) \ .
\ee
This is an {\it ad hoc} procedure that is physically well justified, but could lead to difficulties from a mathematical point of view. Indeed, it is easy to show that the grand potentials $\Xi$ and $\Xi_0$  have different coefficients of quasi-homogeneity. Consequently, the sum $\Xi+\Xi_0$ is not well defined from the point of view of the properties of quasi-homogeneous functions. This is a problem for GTD. In fact, to derive the explicit form of the Legendre invariant metrics (\ref{gI2D}) and (\ref{gII2D}) and the singularity conditions (\ref{singirev})-(\ref{singiiirev}) we have used the Euler identity (\ref{euler}), which explicitly depends on the coefficients of quasi-homogeneity of the function $\Phi(E^a)$.  Therefore, to consider the effect of condensation we are not allowed to use results in which Euler's identity is involved, i.e., the metrics (\ref{gI2D})-(\ref{gII2D}) and the singularity conditions (\ref{singirev})-(\ref{singiiirev}). Consequently, we must use the metric $g^{III}$ given explicitly in Eq.(\ref{gIII2D}), and the thermodynamic potential corresponding to the grand potential $\Xi+\Xi_0$ has to be handled with care. 

However, since we are interested in investigating the condensation in a geometric framework, and the results of GTD do not depend on the choice of thermodynamic potential, we choose the internal energy to study the condensation as a phase transition. In this case, the computations can be carried out in a simple manner \cite{huang}. 
According to (\ref{untd}), the internal energy per particle of a bosonic gas is
\be
u=\frac{U}{N} = \frac{3}{2} k_{_B} T \frac{g_{\frac{5}{2}}(z)}{g_{\frac{3}{2}}(z)} \ .
\label{ubec}
\ee
This equation governs the dynamics of the gas for arbitrary values of $z$. We will therefore use it to investigate the dynamical behavior at the onset of the Bose-Einstein condensation. 

\begin{figure}
\includegraphics[scale=0.3]{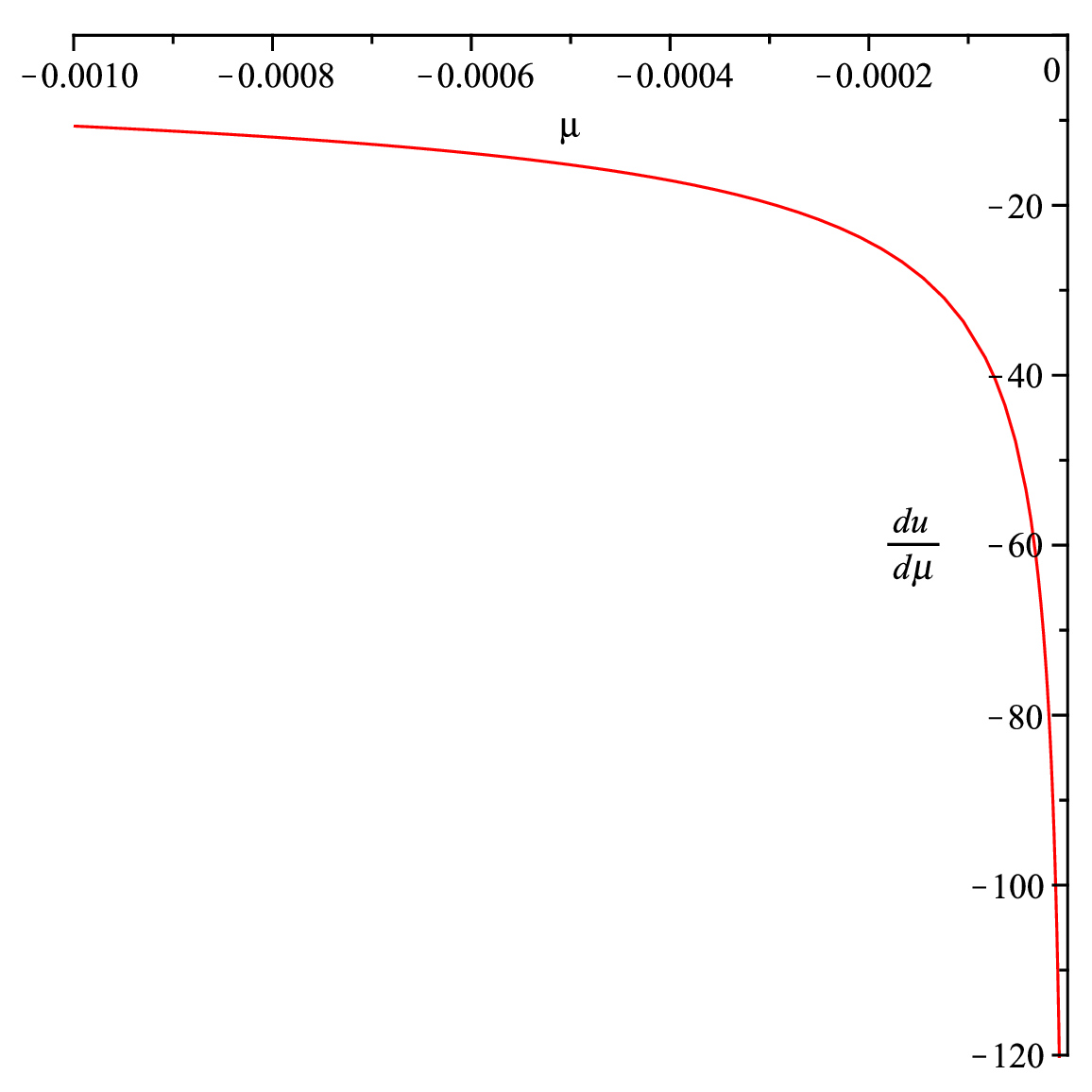}
\caption{Singularity of the first derivative of the thermodynamic potential $u$ with respect to the chemical potential.}  
\label{fig1}
\end{figure}

To apply the Ehrenfest scheme, we compute the derivatives of the thermodynamic potential (\ref{ubec}). Then, it is possible to show that the first derivative with respect to $\mu$ diverges as $\mu\rightarrow 0$. This behavior is illustrated in Fig. \ref{fig1}.
Accordingly, we can conclude that there exists a first order phase transition as the chemical potential approaches zero. 

\begin{figure}
\includegraphics[scale=0.3]{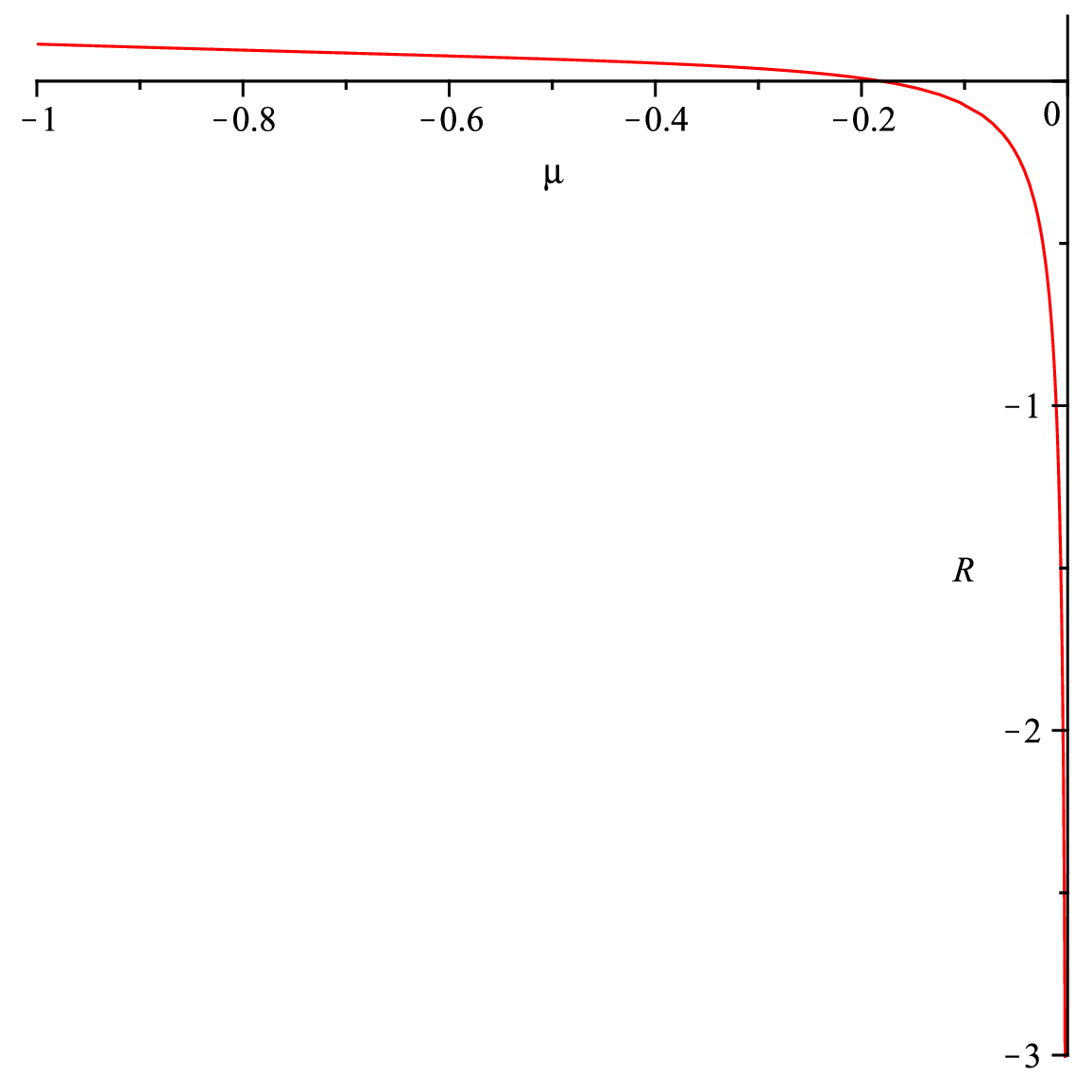}
\caption{Singularity of the curvature scalar near the origin $\mu=0$ for $t=1$. Here we use the value $T_c=1$ for simplicity.}  
\label{fig2}
\end{figure}

We can now investigate the geometric properties of the equilibrium space. It is convenient to rewrite the thermodynamic potential (\ref{ubec}) in terms of the critical temperature. To this end, in the definition of $u$ we consider the maximum particle number, $u=U/N_{max}$, 
to obtain
\be
u = \frac{3}{2\zeta} k_{_B} \frac{T^{5/2}}{T_c^{3/2}}   g_{\frac{5}{2}}(z)   \ ,
\ee
which  for convenience we rewrite as 
\be
u = \frac{3}{2\zeta }\, k_{_B}\, T_c\, t^{5/2}\, g_{\frac{5}{2}} (z) \ , \quad z= e^{\frac{\mu}{k_{_{_B}}T_c t}}\ , \quad t=\frac{T}{T_c} \ .
\ee
This is the fundamental equation that describes the transition of a bosonic gas into a Bose-Einstein condensate in the limit $t\rightarrow 1$ and $\mu\rightarrow 0$.
According to the results of GTD, for the calculation of the metric (\ref{gIII2D}) of the equilibrium space, we identify $u$ with the thermodynamic potential $\Phi$ and $E^a=(t, \mu)$. Then, a straightforward computation leads to the following metric components
\be
g_{tt}= \frac{9}{32  \zeta^2} \left( 5T_c t g_{\frac{5}{2}} - 2 \mu g_{\frac{3}{2}}\right)\left(15 T_c^2 t^2 g_{\frac{5}{2}} -12 T_c t \mu
g_{\frac{3}{2}} + 4 \mu^2 g_{\frac{1}{2}}\right)\ ,
\ee
\be
g_{t\mu}= \frac{45}{32 \zeta^2} t^2 g_{\frac{5}{2}} \left(3 T_c t g_{\frac{3}{2}} - 2\mu g_{\frac{1}{2}} \right)\ ,
\ee
\be
g_{\mu\mu} = \frac{9}{4\zeta^2 T_c} g_{\frac{3}{2}} g_{\frac{1}{2}}\ ,
\ee
where for the sake of simplicity we set $k_{_B}=1$ and drop the argument of the Bose integral $g_n\left(e^{\frac{\mu}{T_c t}}\right)$.
The corresponding curvature tensor can then be computed, but its  final expression cannot be written in a compact form. Nevertheless, we performed a detailed numerical analysis of the behavior of the curvature scalar and found that there is a singularity in the limit $t=1$ and $\mu\rightarrow 0$. Figure \ref{fig2} illustrates this behavior.

This result shows that the values at which the  Bose-Einstein condensation takes place correspond to a curvature singularity of the respective
equilibrium space.


\section{Conclusions}
\label{sec:con}

In the present work, we analyzed the properties of the space of equilibrium states of the ideal quantum gases in the context of geometrothermodynamics. First, we derived the fundamental equations from which we can obtain all the thermodynamic properties of the gases. It was established that a Boltzmann gas, whose components are identical classical particles, possesses a flat equilibrium space, indicating the lack of thermodynamic interaction. This is in accordance with our intuitive interpretation of thermodynamic interaction, which is usually associated with the presence of a potential term in the corresponding Hamiltonian. 

On the other hand, our analysis shows that fermionic and bosonic  quantum gases possess a non-flat equilibrium space, although the corresponding Hamiltonian does not contain a potential term. This result shows that the curvature of the equilibrium space is able to detect the ``effective" thermodynamic interaction, which arises from the quantum nature of the gas components. We interpret this result as an indication in favor of considering the curvature of the equilibrium manifold as a measure of the thermodynamic interaction, instead of the intuitive notion based upon the presence of a potential term in the Hamiltonian. 

We also analyzed the Bose-Einstein condensation from the point of view of GTD. First, we established a particular fundamental equation that allows us to apply the Ehrenfest definition to interpret the Bose-Einstein condensation as a first order phase transition of an ideal bosonic gas. The same fundamental equation is then used to derive the geometric properties of the corresponding equilibrium manifold. We found that the curvature is non-zero, indicating the presence of thermodynamic interaction, and that there exists a curvature singularity at exactly that point in the equilibrium  manifold, where the Bose-Einstein condensation takes place. This
result indicates that GTD is able to correctly describe the thermodynamic properties of ideal classical and quantum gases. 

Ideal quantum gases have been investigated from the point of view information geometric theory 
\cite{amari85,janmru90,ooh99,mirmoh11}. Using the second moments of the energy and particle number fluctuations as the components of a thermodynamic metric, 
in \cite{janmru90}, it was shown that the corresponding curvature can be used as a measure of stability.  For bosons, for example, the curvature tends to zero 
at the classical limit and diverges in the condensation region. The results we obtained in GTD are consistent with this information geometric approach. 
A quantum ideal gas obeying an intermediate 
statistics was investigated in \cite{ooh99}. The thermodynamic curvature is constructed such that it depends on the fugacity 
and the number of particles in a  state, and it turns out to contain information about the stability properties of the system. In the classical limit of a Boltzmann 
gas, however, the curvature does not vanish as expected, but contains contributions from the quantum statistical character of the gas. A different intermediate 
statistics for deformed bosons and deformed fermions was investigated in \cite{mirmoh11}. The corresponding singular points of the curvature were shown to be 
related with condensation even in the case of deformed bosons.

Finally, for the sake of comparison, we used thermodynamic geometry in which the metric of the equilibrium space is given as the Hessian for different thermodynamic potentials. 
In the case of the entropy, the curvature is non-zero with a singularity at a value of $\mu$ that does not coincide with the Bose-Einstein condensate limiting value. If, instead, the internal energy is used as thermodynamic potential the corresponding metric is flat, indicating that no interaction is present. Both results are inconsistent with the thermodynamic properties of ideal quantum gases.


\section*{Acknowledgements} 

We thank R. Paredes and V. Romero-Rochin for stimulating discussions and literature hints and the UNAM-GTD group for helpful comments and discussions. Sasha A. Zaldivar would like to deeply thank Dr. O. Avil\'es-Hern\'andez for his continued support and encouraging and stimulating conversations. She also thanks the UNAM for a graduate scholarship, CVU 493842. This work was partially supported  by UNAM-DGAPA-PAPIIT, Grant No. 114510, and Conacyt-Mexico, Grant No. A1-S-31269.



\end{document}